\documentclass[12pt]{article}   
\usepackage{amsmath}
\usepackage{amssymb}
\usepackage{graphicx}

\begin{document}
\title{de Sitter relativity in central charts}

\author{Ion I. Cot\u aescu 
\thanks{e-mail: i.cotaescu@e-uvt.ro}\\
{West University of Timi\c soara,} \\{V. P\^ arvan Ave. 4, RO-300223, Timi\c soara, Romania}}

\maketitle
\begin{abstract}
The relative geodesic motion in central charts (i.e. static and spherically symmetric) on the $(1+3)$-dimensional de Sitter spacetimes is studied in terms of conserved quantities. The Lorentzian isometries are derived, relating the coordinates of the local chart of a fixed observer  with the coordinates of a mobile chart considered as the rest frame of a massive paticle freely moving on a timelike geodesic. The time dilation and Lorentz contraction are discussed pointing out some notable features of the de Sitter relativity in central charts. 
\end{abstract}


\section{Introduction}
\label{Int}

The simplest $(1+3)$-dimensional spacetimes of special or general relativity are  vacuum solutions of the Einstein equations whose geometry is determined only by the value of the cosmological constant $\Lambda$. These are the Minkowski flat spacetime (with $\Lambda=0$),  and the hyperbolic spacetimes,  de Sitter (dS) with $\Lambda>0$ and Anti-de Sitter (AdS) having $\Lambda<0$. All these spacetimes have highest possible isometries \cite{SW}  representing thus a good framework for studying the role of the conserved quantities with physical meaning \cite{ES,CGRG,CAdS1} in studying the relative geodesic motion.  With their help we constructed recently the dS relativity \cite{CdS} in comoving charts \cite{BD} and the AdS  relativity \cite{CAdS2} in central charts (i. e. static and spherically symmetric) that complete our image about the special relativity in spacetimes with maximal symmetry.

Our approach is based on the idea that the inertial (natural) frames are local charts playing the role of rest frames of massive particles freely moving along timelike geodesics. Moreover, we impose a sinchronisation condition requiring  the origins of the fixed and moving frames to overlap each other at a given time. The conserved quantities on these geodesics help us to mark the different inertial frames whose relative motion can be studied then by using the Nachtmann boosting method of introducing  coordinates in different dS local charts \cite{Nach}.  In this manner we derived the Lorentzian isometries relating  the coordinates of the moving and fixed inertial frames on dS or AdS backgrounds \cite{CdS,CAdS2}.   

The $(1+3)$-dimensional AdS spacetime is the only maximally symmetric spacetime which does not have space translations \cite{SW} since its $\Lambda<0$ produces an attraction of  elastic type such that the geodesic motion is oscillatory around the origins of the central charts  with ellipsoidal closed trajectories. The AdS relativity relates these charts such that, according to the synchronization condition,  the  moving frames may have only rectilinear geodesics whose oscillatory motions are centred in the origin of the fixed frame \cite{CAdS2}. On the contrary,  in the comoving local charts we used so far (i. e. the conformal Euclidean  and de Sitter-Painlev\'e ones), the dS relativity seems to be closer to the Einstein special relativity  since here we have translations and conserved momenta  such that all the geodesic trajectories are rectilinear along the momentum direction  \cite{CAdS2}. 

However,  apart from the comoving charts, the dS spacetime  also has central charts where the geodesic trajectories are no longer rectilinear such that the role of the conserved momentum becomes somewhat obscure. Since the dS relativity in these charts is not yet formulated, we focus here on this problem studying  the role of the conserved quantities along geodesics in describing the relative geodesic motion.   

In order to preserve the coherence of our dS relativity, we use here the same definitions, conventions and initial conditions as in Ref. \cite{CdS}  since then we can take over the results obtained therein without revisiting the entire boosting method which allowed us to construct the dS and AdS relativity. In this manner we obtain a version of the dS relativity in central charts which is perfectly symmetric with the AdS one with respect to the change of the hyperbolic functions into trigonometric ones. 

The principal new results we report here concern  the role of the conserved quantities in determining  the geodesic kinematic in central charts and  the parametrization of the Lorentzian isometries relating fixed and moving central charts. Moreover, we briefly discuss some  notable properties  of these isometries and their consequences upon simple relativistic effects as the time dilation and Lorentz contraction observed in the origins of the fixed and moving frames. 

We start in the second section  with a short review of the central charts where we consider the dS conserved quantities presented in the third section. The next section is devoted to  the timelike geodesics in central charts showing how their  integration constants depend on   the conserved quantities with physical meaning and pointing out the kinematic role of these quantities. In section 5 we solve the relativity problem in central charts deriving  the Lorentzian isometries with different parametrizations. In the last part of this section we discuss the above mentioned simple relativistic effects in the particular case when the measurements are performed in the origins of the mobile and fixed frames. In the last section we present the dS-AdS symmetry which involves all the conserved quantities of  both these spacetimes. 

\section{Central charts on dS spacetimes}
\label{Sec2}

Let us consider the $(1+3)$-dimensional dS spacetime $(M,g)$  which  is a vacuum solution of the Einstein equations with $\Lambda>0$ and positive constant curvature. This, is a  hyperboloid of radius $R=\frac{1}{\omega} = \sqrt{\frac{3}{\Lambda}}$ embedded in the $(1+4)$-dimensional pseudo-Euclidean spacetime $(M^5,\eta^5)$ of Cartesian coordinates $z^A$  (labeled by the indices $A,\,B,...= 0,1,2,3,4$) and metric $\eta^5={\rm diag}(1,-1,-1,-1,-1)$. These coordinates are global, corresponding to the pseudo-orthonormal basis $\{\nu_A\}$ of the frame into consideration, whose unit vectors satisfy $\nu_A\cdot\nu_B=\eta^5_{AB}$. Any point $z\in M^5$ is  represented by the five-dimensional vector $z=\nu_A z^A=(z^0,z^1,z^2,z^3,z^4)^T$ which  transforms linearly under the gauge group $SO(2,3)$  which leave the metric $\eta^5$ invariant.

The local central charts $\{x\}$,  of coordinates $x^{\mu}$ ($\alpha,..\mu,\nu...$ $=$ $0,1,2,3$), can be introduced on $(M,g)$ giving the set of functions $z^A(x)$ which solve the hyperboloid equation,
\begin{equation}\label{hip}
\eta^5_{AB}z^A(x) z^B(x)=-\frac{1}{\omega^2}\,.
\end{equation}
The usual central chart  $\{t,\vec{x}\}$ with Cartesian spaces coordinates $x^i$ ($i,j,k,...=1,2,3$) is defined by
\begin{eqnarray}
z^0(x)&=&\frac{1}{\omega}\chi(r)\sinh(\omega t)\,,\nonumber\\
z^i(x)&=&x^i \,, \label{Zx}\\
z^{4}(x)&=&\frac{1}{\omega}\chi(r)\cosh(\omega t)\,,\nonumber
\end{eqnarray}
where  $r=|\vec{x}|\le \frac{1}{\omega}$ and $\chi(r)=\sqrt{1-\omega {\vec{x}\,}^2}=\sqrt{1-\omega^2 r^2}$. Hereby one obtains the line element
\begin{eqnarray}
ds^{2}&=&\eta^5_{AB}dz^A(x)dz^B(x)\nonumber\\
&=&\chi(r)^2 dt^{2}-\left[\delta_{ij}+\omega^2 \frac{x^{i}x^{j}}{\chi(r)^2}\right]dx^{i}dx^{j}\,.\label{line1}
\end{eqnarray}
The associated central chart $\{t,r,\theta,\phi\}$ with spherical coordinates, canonically related to the Cartesian ones, $\vec{x}\to (r,\theta,\phi)$, has the  line element 
\begin{equation}\label{line2}
ds^2=\chi(r)^2 dt^2-\frac{dr^2}{\chi(r)^2}-r^2(d\theta^2+\sin ^2 \theta\, d\phi^2)\,.
\end{equation}

Apart from the above usual charts, it is useful to consider the central chart $\{\tilde x\}=\{t,\rho,\theta,\phi\}$ resulted after the substitution \cite{P1,P2}
\begin{equation}\label{rrho}
r=\frac{\rho}{\tilde\chi(\rho)}\,, \quad \tilde\chi(\rho)=\sqrt{1+\omega^2\rho^2}\,,
\end{equation}
where $\rho\in [0,\infty)$. Then the embedding equations become
\begin{eqnarray}
z^0(\tilde x)&=&\frac{1}{\omega\tilde\chi(\rho)}\sinh(\omega t)\,,\nonumber\\
z^1(\tilde x)&=&\frac{\rho}{\tilde\chi(\rho)}\sin \theta \cos \phi \,,\nonumber\\
z^2(\tilde x)&=&\frac{\rho}{\tilde\chi(\rho)}\sin \theta \sin \phi \,,\label{Ztx}\\
z^3(\tilde x)&=&\frac{\rho}{\tilde\chi(\rho)}\cos \theta\,,\nonumber\\
z^{4}(\tilde x)&=&\frac{1}{\omega\tilde\chi(\rho)}\cosh(\omega t)\,,\nonumber\\
\end{eqnarray}
while the line element reads
\begin{equation}\label{line3}
ds^2=\frac{1}{\tilde\chi(\rho)^2}\left[ dt^2-\frac{d\rho^2}{\tilde\chi(\rho)^2}-\rho^2(d\theta^2+\sin ^2 \theta\, d\phi^2)\right]\,.
\end{equation}  
In this chart  the components of the four-velocity are denoted as $\tilde u^{\mu}=\frac{d\tilde x^{\mu}}{ds}$.

\section{Conserved quantities}
\label{Sec3}

The dS spacetimes are homegeneous spaces of the gauge group $SO(1,4)$ whose transformations leave invariant the metric $\eta^5$ of the embedding manifold $M^5$ and implicitly Eq. (\ref{hip}).  For this group we adopt the canonical parametrization
\begin{equation}
{\frak g}(\xi)=\exp\left(-\frac{i}{2}\,\xi^{AB}{\frak S}_{AB}\right)\in SO(1,4) 
\end{equation}
with skew-symmetric parameters, $\xi^{AB}=-\xi^{BA}$,  and the covariant generators of the fundamental representation of the $so(1,4)$ algebra carried by $M^5$ having the matrix elements
\begin{equation}
({\frak S}_{AB})^{C\,\cdot}_{\cdot\,D}=i\left(\delta^C_A\, \eta_{BD}^5
-\delta^C_B\, \eta_{AD}^5\right)\,.
\end{equation}
In any local chart $\{x\}$,  defined by the functions $z=z(x)$, each transformation ${\frak g}\in SO(2,3)$ gives rise to the isometry $x\to x'=\phi_{\frak g}(x)$ derived from the system of equations $z[\phi_{\frak g}(x)]={\frak g}z(x)$. 

The  $so(1,4)$ basis-generators with an obvious physical meaning \cite{CGRG} are the energy ${\frak H}=\omega{\frak S}_{04}$, angular momentum  ${\frak J}_k=\frac{1}{2}\varepsilon_{kij}{\frak S}_{ij}$, Lorentz boosts ${\frak K}_i={\frak S}_{0i}$, and the Runge-Lenz-type vector ${\frak R}_i={\frak S}_{i4}$. In addition, we consider the momentum ${\frak P}_i=-\omega({\frak R}_i+{\frak K}_i)$ and its dual ${\frak Q}_i=\omega({\frak K}_i-{\frak R}_i)$ which are nilpotent matrices  of two Abelian three-dimensional subalgebras \cite{CGRG}.

In general, after integrating the geodesic equations, one obtains the geodesic trajectories depending on some integration constants that must get a physical interpretation. This is possible only by expressing them in terms of conserved quantities on geodesics.  
These are given by the Killing vectors associated to the $SO(1,4)$  isometries  which are defined (up to a multiplicative constant)  as \cite{ES},
\begin{equation}
({\frak S}_{AB})\to k_{(AB)\,\mu}=z_A\partial_{\mu} z_B -z_B\partial_{\mu} z_A\,,  
\end{equation}
where $z_A=\eta^5_{AC}z^C$. The principal conserved quantities along a timelike geodesic of a point-like particle of mass $m$ and momentum $\vec{P}$ \cite{CGRG} have the general form  
\begin{equation}
{\cal K}_{(AB)}(x,\vec{P})=\omega k_{(AB)\,\mu}m u^{\mu}
\end{equation}
where $u^{\mu}=\frac{dx^{\mu}(s)}{ds}$ are the components of the covariant four-velocity that satisfy  $u^2=g_{\mu\nu}u^{\mu}u^{\nu}=1$. The conserved quantities with physical meaning \cite{CGRG} are the well-known energy and angular momentum 
\begin{eqnarray}
 E&=&\omega  k_{(04)\,\mu}m u^{\mu}\label{consE}\\
 L_i&=& \frac{1}{2}\varepsilon_{ijk}k_{(jk)\,\mu}m u^{\mu}
\end{eqnarray}
and the $SO(3)$ vectors having the components
\begin{eqnarray}
K_i&=&  k_{(0i)\,\mu} m u^{\mu}\,,\\
R_i&=&  k_{(i4)\,\mu} m u^{\mu}\,,\label{consR}
\end{eqnarray}
and related to  the conserved momentum, $\vec{P}$ and its dual $\vec{Q}$ defined as  \cite{CGRG}
\begin{equation}\label{PQ}
P^i=-\omega(R_i+K_i)\,, \quad Q^i=\omega(K_i-R_i)\,.
\end{equation}
Thus we can construct the five-dimensional matrix 
\begin{equation}
{\cal K}(x,\vec{P})=
\left(
\begin{array}{ccccc}
0&\omega K_1&\omega K_2&\omega K_3&E\\
-\omega K_1&0&\omega L_3&-\omega L_2&\omega R_1\\
-\omega K_2&-\omega L_3&0&\omega L_1&\omega R_2\\
-\omega K_3&\omega L_2&-\omega L_1&0&\omega R_3\\
-E&\omega R_1&-\omega R_2&-\omega R_3&0
\end{array}\right)\,,\label{KK}
\end{equation}
whose elements  transform  as  a  skew-symmetric tensor on $M^5$, according to the rule 
\begin{equation}\label{KAB}
{\cal K}_{(AB)}(x',{\vec{P}}')={\frak g}_{A\,\cdot}^{\cdot\,C}\,{\frak g}_{B\,\cdot}^{\cdot\,D}\,{\cal K}_{(CD)}(x,\vec{P})\,,
\end{equation}
for all ${\frak g}\in SO(1,4)$. Here  ${\frak g}_{A\,\cdot}^{\cdot\,B}=\eta^5_{AC}\,{\frak g}^{C\,\cdot}_{\cdot \,D}\, \eta^{5\,BD}$ are the matrix elements of the adjoint matrix $\overline{\frak g}=\eta^5\,{\frak g}\,\eta^5$. Thus,  Eq. (\ref{KAB}) can be written as  ${\cal K}(x',{\vec{P}}')=\overline{\frak g}\,{\cal K}(x,\vec{P})\,\overline{\frak g}^T$ or simpler, ${\cal K}'=\overline{\frak g}\,{\cal K}\,\overline{\frak g}^T$. 

Notice  that all the conserved quantities carrying space indices ($i,j,...$) transform alike under rotations as $SO(3)$ vectors or tensors. Moreover, the condition  $z^i\propto x^i$ fixes the same (common)  three-dimensional basis $\{\vec{\nu}_1,\vec{\nu}_2,\vec{\nu}_3\}$  in  both the Cartesian charts, of  $M^5$ and respectively $M$.  This means  that the $SO(3)$ symmetry is global \cite{ES} such that we may use the vector notation for the conserved quantities as well as for the local Cartesian coordinates on $M$.
However, this basis must not be confused with that of the local orthogonal frames on $M$. 

For studying these conserved quantities on the timelike geodesics we chose the chart $\{t,\rho,\theta,\phi\}$ taking the angular momentum along the third axis,  $\vec{L}=L\vec{\nu}_3=(0,0,L)$, for restricting the motion in the equatorial plane, with $\theta=\frac{\pi}{2}$  and $\tilde u^{\theta}=0$. Then the non-vanishing conserved quantities can be written as 
\begin{eqnarray}
E&=&\frac{m }{\tilde{\chi}^2}\tilde u^t\,,\label{conE}\\
L&=&\frac{m \rho^2 }{\tilde{\chi}^2}\tilde u^{\phi}\,,\label{conL}\\
K_1&=&\frac{m }{\omega\tilde{\chi}^2}\left(\omega \rho \tilde u^t\cosh\omega t\cos\phi \right. \nonumber\\
&&\left.-\tilde u^{\rho}\sinh\omega t\cos\phi +\rho \tilde u^{\phi}\sinh\omega t\sin\phi\right)\,,
\label{K1}\\
K_2&=&\frac{m }{\omega\tilde{\chi}^2}\left(\omega \rho \tilde u^t\cosh\omega t\sin\phi \right. \nonumber\\
&&\left.-\tilde u^{\rho}\sinh\omega t\sin\phi -\rho \tilde u^{\phi}\sinh\omega t\cos\phi\right)\,,\\
R_1&=&\frac{m }{\omega\tilde{\chi}^2}\left(\omega \rho \tilde u^t\sinh\omega t\cos\phi \right. \nonumber\\
&&\left.-\tilde u^{\rho}\cosh\omega t\cos\phi +\rho \tilde u^{\phi}\cosh\omega t\sin\phi\right)\,,\\
R_2&=&\frac{m }{\omega\tilde{\chi}^2}\left(\omega \rho \tilde u^t\sinh\omega t\sin\phi \right. \nonumber\\
&&\left.-\tilde u^{\rho}\cosh\omega t\sin\phi -\rho \tilde u^{\phi}\cosh\omega t\cos\phi\right)\,,\label{N2}
\end{eqnarray}
while $K_3=R_3=0$. Hereby we deduce the following obvious properties
\begin{equation}\label{property}
\vec{K}\cdot \vec{L}=\vec{R}\cdot \vec{L}=0\,, \quad \vec{K}\land \vec{R}=-\frac{E}{\omega} \vec{L}\,,
\end{equation}
and verify the identity  
\begin{equation}\label{invariant}
E^2-\omega^2\left({\vec{L}\,}^2 +{\vec{R}\,}^2-{\vec{K}\,}^2\right)=m^2\tilde u^2=m^2\,,
\end{equation}
defining the principal invariant corresponding to the first Casimir operator of the $so(1,4)$ algebra. We specify  that in the classical theory the second invariant of this algebra vanishes since there is no spin \cite{CGRG}.

\section{Timelike geodesics}
\label{Sec4}

In the case of the timelike geodesics we may exploit the identity $\tilde u^2=1$ and Eqs. (\ref{conE}) and (\ref{conL}) for obtaining the radial component 
\begin{equation}\label{ur}
\tilde u^{\rho}=\tilde\chi(\rho)^2\left[\frac{E^2}{m^2} \tilde\chi(\rho)^2-\frac{\omega^2 L^2}{m^2}-\frac{L^2}{m^2\rho^2}-1\right]^{\frac{1}{2}}\,,
\end{equation}
that allows us to derive the following prime integrals,
\begin{eqnarray}
\left(\frac{d\rho}{dt}\right)^2-\omega^2\rho^2+\frac{L^2}{E^2\rho^2}&=&1-\frac{\omega^2 L^2}{E^2}-\frac{m^2}{E^2}\,,\\
\frac{d\phi}{dt}&=&\frac{L}{E\rho^2}\,,
\end{eqnarray}
that give the geodesic equations in the plane $(\vec{\nu}_1,\vec{\nu}_2)$ as
\begin{eqnarray}
\rho(t)&=&\left[-\kappa_1+\kappa_2\cosh 2\omega (t-t_0)\right]^{\frac{1}{2}}\,,\label{GEO1}\\
\phi(t)&=&\phi_0+{\rm arctan} \left[\sqrt{\frac{\kappa_2+\kappa_1}{\kappa_2-\kappa_1}}\tanh\omega(t- t_0)\right],\label{GEO2}
\end{eqnarray}
where
\begin{eqnarray}
\kappa_1&=&\frac{E^2-m^2-\omega^2 L^2}{2\omega^2 E^2}\,,\label{Kap1}\\
\kappa_2&=&\frac{1}{2\omega^2 E^2}\left[(E+m)^2+\omega^2 L^2\right]^{\frac{1}{2}} \left[(E-m)^2+\omega^2 L^2\right]^{\frac{1}{2}}\,,\nonumber\\
\label{Kap2}
\end{eqnarray}
satisfy the identity 
\begin{equation}\label{Kap3}
\kappa_2^2-\kappa_1^2=\frac{L^2}{\omega^2 E^2}\,.
\end{equation}
Thus we solve the geodesic equation in terms of conserved quantities which give a physical meaning to the principal integration constants. The remaining  ones,  $t_0$ and $\phi_0$, determine only the initial position of the mobile and implicitly of its trajectory. 

In this manner, we recover the well-known behavior of the time-like geodesic  trajectory which is a hyperbola that for $\phi_0=0$ can be written easily the in Cartesian space coordinates $(\rho,\phi)\to (\tilde x^1,\tilde x^2)$ of the plane   $(\vec{\nu}_1,\vec{\nu}_2)$ as  
\begin{eqnarray}
\tilde x^1(t)&=&\rho(t)\cos\phi(t)=\rho_{-}\cosh\omega(t-t0)\,,\\
\tilde x^2(t)&=&\rho(t)\sin\phi(t)=\rho_{+}\sinh\omega(t-t0)\,,
\end{eqnarray}
where
\begin{equation}\label{rhomm}
\rho_{-}=\sqrt{\kappa_2-\kappa_1}\,, \quad \rho_{+}=\sqrt{\kappa_2+\kappa_1}\,.
\end{equation}
In the usual Cartesian coordinates these equations read
\begin{eqnarray}
x^1(t)&=&\frac{\rho_{-} }{\sqrt{1+\omega^2 \rho(t)^2}}\,\cosh \omega (t-t_0)\,,\label{geo1} \\
x^2(t)&=&\frac{\rho_{+} }{\sqrt{1+\omega^2 \rho(t)^2}}\,\sinh \omega (t-t_0)\,,\label{geo2}\\
x^3(t)&=&0\,,\label{geo3}
\end{eqnarray}
as it results from Eqs. (\ref{rrho}), (\ref{GEO1}) and (\ref{GEO2}).

{ \begin{figure}
    \centering
    \includegraphics[scale=.56]{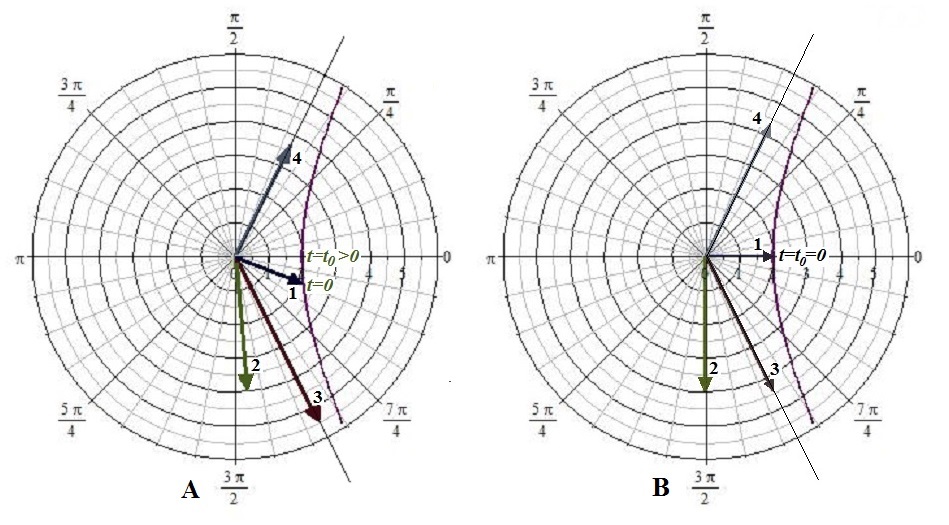}
    \caption{de Sitter timelike geodesics with  $\phi_0=0$ and $t_0>0$ (panel A) or $t_0=0$ (panel B).  The marked vectors are:  $1:\, \frac{\vec{K}}{E}$, $2:\, \frac{\vec{R}}{E}$, $3:\, -\frac{\vec{P}}{\omega E}$ and $4:\,\frac{\vec{Q}}{\omega E}$}
  \end{figure}}
  
Now it remains to analyze the role of the conserved vectors $\vec{K}$ and $\vec{R}$ that depends on $E$ and $L$ as well as on $t_0$ and $\phi_0$. In the simpler case of $\phi_0=0$ their non-vanishing components read
\begin{eqnarray}
K_1&=&E\rho_{-}\cosh\omega t_0\,, \label{KER1}\\
K_2&=&-E\rho_{+}\sinh\omega t_0\,,\label{KER2}\\
R_1&=&E\rho_{-}\sinh\omega t_0 \,,\label{NER1}\\
R_2&=&-E\rho_{+}\cosh\omega t_0\,,\label{NER2}
\end{eqnarray}
complying with the specific properties
\begin{eqnarray}
K=|\vec{K}|&=&E \left[-\kappa_1+\kappa_2\cosh 2\omega t_0\right]^{\frac{1}{2}}\,,\label{prop1}\\
R=|\vec{R}|&=&E \left[\kappa_1+\kappa_2\cosh 2\omega t_0\right]^{\frac{1}{2}}\,,\\
\vec{K}\cdot \vec{R}&=&E^2\kappa_2 \sinh 2\omega t_0\,.\label{prop3}
\end{eqnarray}
Moreover, the conserved momentum and its dual defined by Eq. (\ref{PQ}) have the norms 
\begin{eqnarray}
P&=&|\vec{P}|=\sqrt{2\kappa_2}\,\omega E e^{\omega t_0}\,,\label{norPQ}\\
 Q&=&|\vec{Q}|=\sqrt{2\kappa_2}\,\omega E e^{-\omega t_0}\,,\label{norPQ1}
\end{eqnarray}
and satisfy
\begin{equation}
\vec{P}\cdot\vec{Q}=-2 \kappa_1 \omega^2 E^2\,.
\end{equation}
Hereby we understand that the vector $\frac{\vec{K}}{E}$  indicates the position of the particle of mass $m$ at $t=0$ (Fig. 1A) such that for $t_0=0$ this vector  lays over the semi major axis being orthogonal on  $\frac{\vec{R}}{E}$  (Fig. 1B). It is remarkable that for any $t_0$ the vector $\vec{P}$ is oriented along the lower asymptote  while $\vec{Q}$ gives the direction of the upper one (as in Figs. 1A and 1B). Thus the vector $\vec{Q}$, whose role in comoving charts was rather unclear  \cite{CdS}, gets now a precised  physical meaning.

All these properties are independent on the value of  $\phi_0$ which gives only the rotation of the major axis in the plane  $(\vec{\nu}_1,\vec{\nu}_2)$. Nevertheless, in the Appendix A we give the general form of all these conserved quantities  calculated for an arbitrary $\phi_0\not=0$.  

An important particular case is when the geodesic is passing through the origin  since then the trajectory is rectilinear with  ${L}=0$ and $\kappa_{-}=0$. Consequently,  the vectors $\vec{P}$ and $\vec{Q}$ become parallel having the norms 
\begin{equation}
P=e^{\omega t0}\sqrt{E^2-m^2}\,, \quad Q=e^{-\omega t0}\sqrt{E^2-m^2}\,,
\end{equation}
resulted from Eqs.(\ref{norPQ}) and (\ref{norPQ1}), and we obtain the geodesic equation
\begin{equation}\label{geolin}
{x}^i(t)=\frac{1}{\omega}\frac{{P}^ie^{-\omega t_0}\sinh\omega(t-t_0)}{\sqrt{E^2+P^2e^{-2\omega t_0}\sinh^2\omega(t-t_0)}}\,,
\end{equation} 
and the four-velocity
\begin{eqnarray}
u^0&=&\frac{1}{Em}\left[E^2+P^2 e^{-\omega t_0}\sinh^2 \omega(t-t_0)\right]\,,\label{velo1}\\
u^i&=&\frac{E}{m}\frac{P^i e^{-\omega t_0}\cosh \omega(t-t_0)}{\sqrt{E^2+P^2 e^{-2\omega t_0}\sinh^2 \omega(t-t_0)}}\,,\label{velo2}
\end{eqnarray}
Notice that if, in addition, we take $t_0=0$ then we have $\vec{K}=0$,  $\vec{P}=\vec{Q}=-\omega\vec{R}$ and $P=Q=\sqrt{E^2-m^2}$, as in special relativity.

\section{Relativity}
\label{Sec5}

Recently we have studied the relativitive geodesic motion on dS \cite{CdS} and AdS manifolds \cite{CAdS2}, applying the Nachtmann method  of boosting coordinates \cite{Nach}.  In the case of the AdS spacetimes we used central charts while for the dS spacetimes we considered only comoving charts (i. e. the conformal Euclidean and de Sitter-Painlev\'e ones) where the geo\-desics are rectiliniar \cite{CdS}. Here we complete this study constructing the dS relativity in central charts by taking over the results obtained previously in comoving charts, without revisiting the entire boosting method. 

\subsection{Lorentzian isometries}
\label{Sec51}

The problem of the relative motion is to find how an arbitrary geodesic trajectory and the corresponding conserved quantities can be measured by different observers. The local charts may play the role of inertial frames related through isometries.  Each observer has its own proper frame $\{x\}$ in which he stays at rest in origin on the world line along the vector field $\partial_t$ \cite{CdS}. Here we are interested by the inertial frames defined as  proper frames of massive particles freely moving along geodesics. Then each mobile inertial frame can be labeled by the conserved quantities determining the geodesic of the carrier particle which stays at rest in its origin \cite{CdS}.

In what follows we consider two observers assuming that the first one, $O$, is fixed in the origin  of his proper frame $\{x\}$ observing what happens in a mobile frame $\{x'\}$ of the observer $O'$ which is simultaneously the proper frame of  $O'$ and of a carrier particle of mass $m$ moving along a timelike geodesic with given parameters. This relativity does make sense only if we can compare the measurements of these observers imposing the synchronization condition of their clocks. This means that, at a given common initial time, the origins of these frames must coincide.  However, this condition is restrictive since this forces the geodesic of the  particle carrying the mobile frame  to across the origin of the fixed frame $O$. Consequently,  its trajectory is rectilinear (with $\vec{L}=0$)  in a given direction determined by its conserved momentum $\vec{P}$ as in Eq. (\ref{geolin}).  

The choice of the synchronization condition is a delicate point since the form of the isometry relating the fixed and mobile frames, called Lorenzian isometry, is strongly dependent  on this condition. For this reason we use the same condition as in the case of the comoving frames \cite{CdS} since then the Lorenzian isometry is generated by the same transformation of the $SO(1,4)$ group. Therefore, we set the synchronization condition at $t=t'=0$ when $\vec{x}(0)={\vec{x}}'(0)=0$ such that the origins of the both frames, $O$ and $O'$, overlap the point $z_o=(0,0,0,0,{\omega}^{-1})^T \in M^5$ which was the fixed point in constructing the dS manifold as the space of left cosets $SO(1,4)/ L^{\uparrow}_{+}$  where the Lorentz group $L^{\uparrow}_{+}$ is the stable group of $z_o$ \cite{CdS}. 

Under such circumstances, the synchronization condition is the same as in the case of the comoving charts and we can take over the $SO(1,4)$ transformation  generating the Lorentzian isometry between the frames $O'$ and $O$. This has the form \cite{CdS}
\begin{eqnarray}
{\frak g}(\vec{P})&=&\exp\left(-i {P}^i {\frak K}_i\,\frac{1}{P}\,{\rm arcsinh}\frac{P}{m}\right)\nonumber\\
&=&
\left(
\begin{array}{ccccc}
\frac{E}{m}&\frac{P^1}{m}&\frac{ P^2}{m}&\frac{ P^3}{m}&0\\
\frac{P^1}{m}&1+{n_p^1}^2\nu&n_p^1 n_p^2\nu&n_p^1 n_p^3\nu&0\\
\frac{P^2}{m}&n_p^1 n_p^2\nu&1+{n_p^2}^2\nu&n_p^2 n_p^3\nu&0\\
\frac{ P^3}{m}&n_p^1 n_p^3\nu&n_p^2 n_p^3\nu&1+{n_p^3}^2\nu&0\\
0&0&0&0&1
\end{array}\right)\,,\label{Lorbust}
\end{eqnarray}
where $\vec{n}_P=\frac{\vec P}{P}$ and $\nu=\left(\frac{E}{m}-1\right)$. The four-dimensional restriction of this transformation is a genuine Lorentz boost such that  
 ${\frak g}(\vec{P})^{-1}={\frak g}(-\vec{P})$ and ${\frak g}(0)={\frak e}$. This transformation generates the Lorenzian isometry  and transforms the conserved quantities according to Eq. (\ref{KAB}).

The direct  Lorenzian isometry, $x=\phi_{{\frak g}(\vec{P})}(x')$, between the coordinates of the mobile and fixed frames, results from the system of equations $z(x)={{\frak g}(\vec{P})}z(x')$  as 
\begin{eqnarray}
t(t',{\vec{x}\,}')&=&\frac{1}{\omega}\,{\rm arctanh}\left(\frac{E}{m}\tanh\omega t'\right. \nonumber\\
&&\left.+\frac{\omega}{m}\,\frac{{\vec{x}\,}'\cdot \vec{P}}{\sqrt{1-\omega^2 |\vec{x}\,'|^2}}\, {\rm sech}\,\omega t'\  \right)\,,\label{Lort}\\
\vec{x}(t',{\vec{x}\,}')&=&{\vec{x}\,}'+\frac{\vec{P}}{m}\left(\frac{{\vec{x}\,}'\cdot\vec{P}}{E+m}\right.\nonumber\\
&&\left.+\frac{1}{\omega}\sqrt{1-\omega^2 |\vec{x}\,'|^2}\sinh\omega t'\right)\,,\label{Lox}
\end{eqnarray}
while the inverse one has to be obtained by changing $x \leftrightarrow x'$ and $\vec{P}\to -\vec{P}$.    Obviously, in the limit of $\omega \to 0$ we recover the usual Lorentz transformations of special relativity. 

We verify  first that the geodesic trajectory of the carrier particle can be recovered  from the parametric equations in $t'$ obtained by substituting ${\vec{x}\,}'=0$ in Eqs. (\ref{Lort}) and (\ref{Lox}). Then  we obtain the trajectory of the origin $O'$ denoted as
\begin{equation}
{x}^i_*(t)=\frac{{P}^i\sinh(\omega t)}{\omega \sqrt{E^2+P^2\sinh^2\omega t}}\,,
\end{equation}
which is just Eq. (\ref{geolin}) with $t_0=0$, corresponding to our initial condition $\vec{x}_*(0)=0$.   The components of the four-velocity are those given by Eqs. (\ref{velo1}) and (\ref{velo2}) for $t_0=0$ when we have  $E=mu_*^0(0)$ and  ${P}^i=m {u}^i_*(0)$.  This means that $E$ and ${P}^i$ are the components of the energy-momentum four-vector of the carrier particle when this is passing through the origin of the fixed frame. 

This suggests us to consider as principal parameter the velocity of the carrier particle at $t=0$, defined usually as $\vec{V}=\frac{\vec{P}}{E}$. Then we may put the above formulas in  forms closer to those of special relativity, eliminating the mass $m$ of the carrier particle.  This can be done by changing the parametrization of ${\frak g}(\vec{P})$  seting,  
\begin{equation}
{E}=\gamma m\,,\quad \vec{P}=\gamma m\vec{V}\,,\quad  \gamma=\frac{1}{\sqrt{1-V^2}}\,,
\end{equation}  
such that  we can rewrite
\begin{equation}\label{gfin1}
{\frak g}(\vec{P}) \to{\frak g}(\vec{V})=\exp\left( -i{V}^i{\frak K}_i\,\frac{1}{V}\, {\rm arctanh}\left(V\right)   \right)\,,
\end{equation}
obtaining the new expression of the Lorentzian isometry 
\begin{eqnarray}
t(t',{\vec{x}\,}')&=&\frac{1}{\omega}\,{\rm arctanh}\,\gamma \left(\tanh\omega t'\right.\nonumber\\
&&\left.+\,\frac{\omega{\vec{x}\,}'\cdot \vec{V}}{\sqrt{1-\omega^2 |\vec{x}\,'|^2}}\,{ \rm sech}\,\omega t'\  \right)\,,\label{Lort1}\\
\vec{x}(t',{\vec{x}\,}')&=&{\vec{x}\,}'+\gamma \vec{V}\left({\vec{x}\,}'\cdot\vec{V}\frac{\gamma}{1+\gamma}\right.\nonumber\\
&&\left.+\frac{1}{\omega}\sqrt{1-\omega^2 |\vec{x}\,'|^2}\sinh\omega t'\right)\,,\label{Lox1}
\end{eqnarray}
that may be used in applications. 

The transformations ${\frak g}(\vec{V})$ generating these isometries, transform simultaneously all the conserved quantities. If those of the mobile frame are encapsulated in the matrix ${\cal K}'$ as in Eq. (\ref{KK}), then the corresponding ones measured in the fixed frame are the matrix elements of the matrix
\begin{equation}\label{KKK}
{\cal K}=\overline{\frak g}(\vec{V}) \,{\cal K}'\, \overline{\frak g}(\vec{V})^T\,.
\end{equation}
Thus we obtain the principal tools in studying the relative motion in central charts on dS spacetimes. 

\subsection{Simple relativistic effects}
\label{Sec52}  

The principal feature of the Lorenzian isometries in dS static charts is that the domains  of these mappings do not span the entire static charts involved in such transformations.   Indeed, the condition $|\tanh z|\le 1\,, \forall z\in {\Bbb R}$ indicates that Eq. (\ref{Lort1}) does make sense only in the domain ${\cal D}'$ where the function
\begin{equation}\label{Bfa}
B(\vec{x}')=\omega\frac{ \vec{V}\cdot{\vec x}'}{\chi({\vec x}')}=\omega\frac{ \vec{V}\cdot{\vec x}'}{\sqrt(1-\omega^2|{\vec x}'|^2)}\,.
\end{equation}
satisfies
\begin{equation}\label{resA1}
-\frac{1}{\gamma}\cosh\omega t'-\sinh\omega t'\le B(\vec{x}')\le \frac{1}{\gamma}\cosh\omega t'-\sinh\omega t'\,,
\end{equation}
restricting thus domain of the coordinates $(t', \vec{x}')$ of the mobile frame. 
This condition determines the field of view of the observer $O$ and guarantees that after this transformation we obtain well-defined Cartesian coordinates that satisfy the condition $|\vec{x}(t',\vec{x}')|\le \frac{1}{\omega}$ imposed by the existence of the cosmological horizon. For the inverse Lorentzian isometry, we obtain the similar condition   defining the domain ${\cal D}$ of this transformation. 

It remains to investigate how this restriction works determining the domain ${\cal D}'$.  Assuming that the observation is along an arbitrary direction we denote $\alpha={\rm angle}(\vec{x}',\vec{V})$ such that  we can write $B(\vec{x}')=\omega V  \rho(\vec{x}')\cos \alpha$. Here $\rho$ is the radial coordinate defined by Eq. (\ref{rrho}) that is free of any restriction, taking values in the domain $[0,\infty)$.  Then, the condition (\ref{resA1}) restricts the observation at the points  $(t',{\vec x}')$ which  satisfy $| \rho(\vec{x}')\cos\alpha|\le  \rho_{lim}(V,t')$ where the function 
\begin{equation}
\rho_{lim}(V,t')=\frac{1}{\omega V}\left(\frac{1}{\gamma}\cosh\omega t'-{\rm sign}(t')\sinh\omega t' \right)
\end{equation} 
is positively defined on the domain $[-t'_m,t'_m]$, vanishing  for  $t'=\pm t'_m$ where $t'_m=\frac{1}{\omega}{\rm arctanh} \frac{1}{\gamma}$.  Hence, we may conclude that Eq. (\ref{resA1}) gives rise to non-trivial restrictions that seem to be specific for the static charts. 

More interesting are the simple relativistic effects as  the time dilation (observed in the twin paradox) and the Lorentz contraction. In general, these effects are quite complicated since they are strongly dependent on the position where the time and length are measured.  Let us convince this assuming that the measurement is performed in the point $A$ of arbitrary position vector $\vec{a}$,  fixed rigidly to the mobile frame $O'$. Then we may  write the general relations 
\begin{eqnarray}
\delta t &=&\left.\frac{\partial t(t',{\vec{x}}')}{\partial t'}\right| _{{\vec{x}}'=\vec{a}}\delta t' + \left.\frac{\partial t(t',{\vec{x}}')}{\partial x^{\prime\, i}}\right| _{{\vec{x}}'=\vec{a}}\delta  x^{\prime\, i}\,,\\
\delta x^j &=&\left.\frac{\partial x^j(t',{\vec{x}}')}{\partial t'}\right| _{{\vec{x}}'=\vec{a}}\delta t' +\left.\frac{\partial x^j(t',{\vec{x}}')}{\partial x^{\prime\, i}}\right| _{{\vec{x}}'=\vec{a}}\delta  x^{\prime\, i}\,,
\end{eqnarray} 
allowing us to relate among themselves the quantities $\delta t, \delta x^j$ and $\delta t',\delta x^{\prime\, j}$  measured by the observers $O$ and respectively $O'$.   

We consider first a clock in $A$  indicating $\delta t'$ without changing its position such that $\delta x^{\prime\, i}=0$. Then, after a little calculation, we obtain the time dilation observed by $O$,  $\delta t(t)=\delta t'\, \tilde\gamma(t)$,  given by the function
\begin{equation}\label{gamt}
\tilde\gamma(t)=\gamma\left[ 1-B(\vec{a})\sinh \omega t'(t)\right]\frac{\cosh^2 \omega t}{\cosh^2\omega t'(t)}\,,
\end{equation} 
which depends on the parameter $B(\vec{a})$ defined by Eq. (\ref{Bfa}) for $\vec{x}'=\vec{a}$ and the function 
\begin{eqnarray}\label{dtfin}
t'(t)&=&\frac{1}{\omega}{\rm arctanh}\left[  \frac{1}{\gamma(B({\vec a})^2+1)}\left(\tanh \omega t\right.\right.\nonumber\\
&&+\left. \left.B({\vec a})\sqrt{\gamma^2 B({\vec a})^2+\gamma^2-1+{\rm sech}^2\omega t} \right) \right]
\end{eqnarray}
resulted after inverting Eq. (\ref{Lort1}) with $\vec{x}'=\vec{a}$. Similarly but with the supplemental simultaneity condition $\delta t=0$ we derive the Lorentz contraction of an arbitrary  $\delta\vec{x}'$ that reads
\begin{eqnarray}\label{dxfin}
\delta {\vec x}(t)&=&\gamma\, \vec{n}_V\cdot \delta\vec{x}' -B({\vec a})\, \vec{a}\cdot \delta\vec{x}'\sinh \omega t'(t) \nonumber\\
&&+\frac{\gamma^2-1}{\gamma}\frac{\cosh^2 \omega t'(t)}{B(\vec{a})\sinh \omega t'(t)-1}\nonumber\\
&&\times \left[\vec{V}\cdot \delta\vec{x}'+\frac{(\vec{V}\cdot\vec{a})(\vec{a}\cdot \delta\vec{x}')}{\chi(\vec{a})^2}\right]
\end{eqnarray}
where $\vec{n}_V=\frac{\vec{V}}{V}$.

\begin{figure}
\begin{minipage}{\columnwidth}
\centering
\includegraphics[scale=0.65]{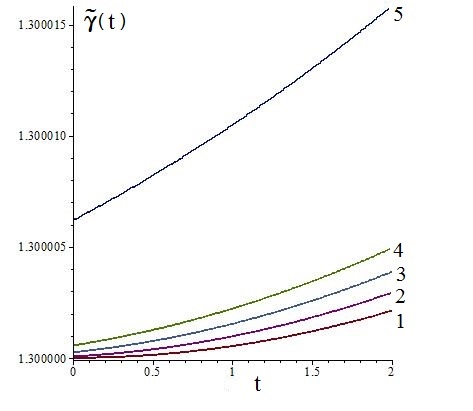}
\end{minipage}
\caption{Functions $\tilde\gamma(t)$ for $\omega=0.01$, $V=0.2$ and $a=0$ (1), $a=1$ (2), $a=2$ (3), $a=3$ (4) and $a=6$ (5).}
\label{fig.1}
\end{figure}

Thus we obtained the general formulas of the simple relativistic effects which hide a large phenomenology that cannot be exhausted here because of the difficulties in studying  analytically the general case of an arbitrary $\vec{a}$. For this reason we  restrict ourselves to the case of  $\vec{a}\cdot \vec{V}=0$ assuming that $\delta\vec{x}'$ is parallel with $\vec{V}$. Then, by taking $B(\vec{a})=0$ in Eqs.(\ref{gamt}), (\ref{dtfin}) and (\ref{dxfin}) we obtain
\begin{equation}
\delta t=\delta t'\, \tilde\gamma(t) \,, \quad \delta x_{\parallel}=\delta x'_{\parallel}\frac{1}{\tilde\gamma(t)}
\end{equation} 
where the function $\tilde\gamma(t)$ takes the simple form
\begin{eqnarray}
\tilde\gamma(t)&=&\frac{\gamma}{\cosh^2 \omega t'(t)-\gamma^2\sinh^2\omega t'(t)}\nonumber\\
&=&\gamma\cosh^2\omega t - \frac{1}{\gamma} \sinh^2\omega t\,,\label{gamt1}
\end{eqnarray}
since now   $\gamma\, {\rm than}\,\omega t'(t)= {\rm than}\,\omega t$. Hereby we recover again the well-known condition of the flat case,   $\delta t \delta x_{||}=\delta t' \delta x'_{||}$, that also  holds in the dS relativity in comoving charts \cite{CdS}.

The function $\tilde\gamma(t)$ is defined on the  domain $(-\infty, \infty)$  taking values in the codomain $[\gamma, \infty)$.  The time dilation observed by $O$  increases to infinity as
\begin{equation}
\tilde\gamma(t)\sim \frac{1}{4}\left(\gamma -\frac{1}{\gamma}\right)e^{2\omega t}
\end{equation}
when $t\to \infty$  since the observer $O$  sees how the clock in $O'$ lats more and more such that $t$ tends to infinity when  $t'$ is approaching to $t_m$.  Notice that the observer $O'$ measures the same dilation of the time $t'$ of a clock staying  at rest in $O$. 

In general, for the clocks situated in arbitrary space points the problem  is much more complicated and cannot be solved  without resorting to numerical method. As an example,   we present in Fig. 2 the functions $\tilde\gamma(t)$ for different norms $a=|\vec{a}|$ of the position vector $\vec{a}$ oriented parallel with $\vec{V}$. Other  interesting and attractive conjectures may be studied numerically starting with the above presented approach.

\section{Remark on the dS-AdS symmetry}
\label{Sec6}

The Lorentzian isometry given by Eqs. (\ref{Lort}) and (\ref{Lox}) is related to the corresponding AdS isometry \cite{CAdS2} through the transformation $\omega\to i\omega$.  This is the effect of the well-known dS-AdS symmetry under this transformation, arising when one uses the same type of local charts in both these spacetimes. Moreover, we must specify that this symmetry is general since the dS conserved quantities  transform into AdS ones  as    

\begin{tabular}{llcl}
\hspace*{5mm}&dS \cite{CdS}&$\omega\to i\omega$&AdS \cite{CAdS2}\\ 
&$E$&$\to$&$E$\\
&$\vec{L}$&$\to$&$\vec{L}$\\
&$\vec{K}$&$\to$&$\vec{K}$\\
&$\vec{R}$&$\to$&$i\vec{N}$\\
&$\vec{P}=-\omega(\vec{K}+\vec{R})$&$\to$&$\omega(\vec{N}-i \vec{K})$\\
&$\vec{Q}=\omega(\vec{K}-\vec{R})$&$\to$&$\omega(\vec{N}+i \vec{K})$
\end{tabular}\\
regardless the local charts we consider. In other respects, this explains why  in AdS spacetimes we do not have a real-valued conserved momentum. 

Concluding we can say that the dS relativity in the conformal charts is closer to the Einstein special relativity having only rectilinear geodesics along the momentum directions, while in static charts the dS relativity is symmetric with the AdS one. Obviously, in the flat limit (when $\omega\to 0$) the dS and AdS relativity tend to the usual special relativity in Minkowski spacetime \cite{CGRG,CAdS1}.   

Finally we note that this symmetry also holds at the level of the quantum theory  where the quantum observables are conserved operators corresponding to the conserved quantities considered above,  having the same physical meaning \cite{ES,CGRG}. We remind the reader that in Ref. \cite{ES} the conserved observables of the covariant quantum fields of any spin on dS and AdS backgrounds are derived laying out the dS-AdS symmetry. However, now it is premature to  discuss how this symmetry may be extended to the quantum field theory  since, even on dS spacetimes we have already the QED in Coulomb gauge \cite{CQED},  on AdS spacetimes a similar theory was not yet constructed.

\appendix

\section{Inverse problem}

There are situations when we know the integration constants $\kappa_1$, $\kappa_2$, $\phi_0$ and $t_0$ and we need to find the physical conserved quantities. Then from Eqs. (\ref{Kap1}) and (\ref{Kap3}) we deduce
\begin{eqnarray}
E&=&\frac{m}{\sqrt{(\kappa_1\omega^2-1)^2-\kappa_2^2\omega^4}}\,,\\
L&=&\frac{m\omega\sqrt{\kappa_2^2-\kappa_1^2}}{\sqrt{(\kappa_1\omega^2-1)^2-\kappa_2^2\omega^4}}\,.
\end{eqnarray}  
Furthermore, from Eqs. (\ref{K1})-(\ref{N2}), calculated in aphelion, at $t=t_0$, where $\tilde u^{\rho}=0$ and $\tilde u^t$ and $\tilde u^{\phi}$ result from Eqs. (\ref{conE}) and (\ref{conL}), we derive the non-vanishing components
\begin{eqnarray}
K_1&=&E\rho_{-}\cos\phi_0\cosh\omega t_0+E\rho_{+}\sin\phi_0\sinh\omega t_0\,,\\
K_2&=&E\rho_{-}\sin\phi_0\cosh\omega t_0-E\rho_{+}\cos\phi_0\sinh\omega t_0\\
R_1&=&-E\rho_{-}\cos\phi_0\sinh\omega t_0+E\rho_{+}\sin\phi_0\cosh\omega t_0\,,\\
R_2&=&-E\rho_{-}\sin\phi_0\sinh\omega t_0-E\rho_{+}\cos\phi_0\cosh\omega t_0\,.
\end{eqnarray}
while $K_3=N_3=0$. These components satisfy the properties  (\ref{prop1})-(\ref{prop3}).

\end{document}